\documentclass[superscriptaddress,
preprintnumbers,amsmath,amssymb]{revtex4}
\usepackage{graphicx}% Include Figure files
\usepackage{dcolumn}% Align table columns on decimal point
\usepackage{bm}% bold math
\usepackage{latexsym}
\begin{document}
\title{Alice in a micro-factory: modeling materials and mechanisms\\ of natural nano-machines{\footnote{A popularized version of the introductory lecture in an undergraduate course on ``Natural Nano-machines'' taught at IIT, Kanpur}}}
\author{Debashish Chowdhury}
\affiliation{Department of Physics\\
Indian Institute of Technology\\
Kanpur 208016, India.
}
\date{\today}
\begin{abstract}
Imagine an under water factory which is about $10 \mu$m long in each
direction. The factory is filled with machines, each typically a few
tens of nanometers long, which perform specific tasks and operate in a
well coordinated manner. A cell, the structural and functional unit
of life, is not very different from this micro-factory. In this 
article, I begin with Alice's guided tour of this micro-factory in 
her dream during which the guide shows her wide varieties of the 
nano-machines in this factory. The style of presentation of the first 
half of this article is inspired by Lewis Carrol's {\it Alice in 
Wonderland}. In the second half, I introduce the methods of studying 
the materials and mechanisms of the molecular machines through 
dialogues; the three participants in this discussion are Alice, her 
elder brother Alex and her father Albert. The style of presentation 
of the second half of this article, in terms of a dialogue, is adapted 
from Galileo's {\it Dialogue Concerning the Two Chief World Systems}. 
Albert, a professor of biophysics, emphasizes the crucial differences 
between the mechanisms of the natural nano-machines and those of their 
macroscopic counterparts. He also points out some practical applications 
of this interdisciplinary research in biomedical science and nano-technology.
\end{abstract}
\maketitle

\section{Introduction}

Alice was sitting on a window seat beside her father inside the aircraft. 
The departure was getting delayed because of the bad weather. She was 
coming back to Delhi with her parents after a two-week vacation with her 
grand parents. Soon her final year in school will begin. Her father is a 
professor of biophysics at a university in Delhi and he had utilized the 
vacation to prepare his lecture notes for the next semester. Alice picked 
up the note book from her father's tray table and started reading the 
following paragraph from the notes.

``Three centuries ago Marcello Malpighi conjectured ``Nature, in order 
to carry out the marvelous operations in animals and plants, has been 
pleased to construct their organized bodies with a very large number 
of machines, which are of necessity made up of extremely minute parts 
so shaped and situated, such as to form a marvelous organ, the 
composition of which are usually invisible to the naked eye, without 
the aid of the microscope.'' (As quoted by Marco Piccolino \cite{picco})''.

%%%%%%%%%%%%%%%%%%%%%%%%%%%%%%%%%%%%%%%%%%%%%%%%%%%%%%%%%%%%%%
\begin{figure}[h]
\begin{center}
\includegraphics[width=0.5\columnwidth]{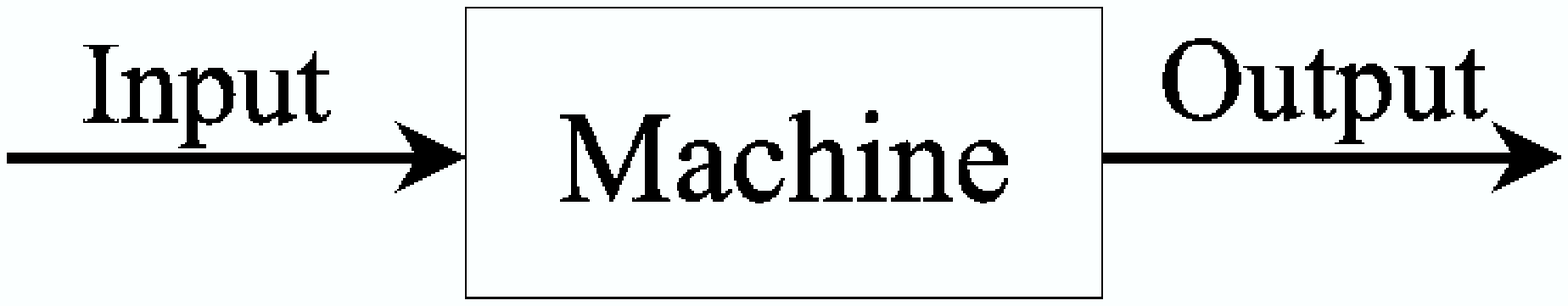}
\end{center}
\caption{A schematic black-box like representation of a machine; 
only the input and the output are mentioned explicitly.
}
\label{fig-machine}
\end{figure}
%%%%%%%%%%%%%%%%%%%%%%%%%%%%%%%%%%%%%%%%%%%%%%%%%%%%%%%%%%%%%

Alice had already learnt a lot about machines from her physics teacher 
in school. A machine is a device, having a unique purpose, that augments 
or replaces human or animal effort for the accomplishment of physical 
tasks. The operation of a machine may involve the transformation of 
chemical, thermal, electrical, or nuclear energy into mechanical energy 
or vice versa.  All machines have an {\it input}, an {\it output}, and 
a {\it transforming} (or, modifying) and a {\it transmitting} device  
(see fig.\ref{fig-machine}).
A man-made complex macroscopic device is usually an assembly of simpler 
components, each of which is designed to achieve a specific function. 
For example, a hair dryer consists of a heater, a fan and a switch 
suitably connected by electric wires and assembled in a compact 
framework; combined and coordinated operation of these components 
gives rise to the function of the hairdryer- blowing hot wind. 

Alice continued with the next paragraph in her father's note book.
``Naturally, to understand how a machine works, one has to understand  
the structure and functions of its components. So, the belief in  
the role of organic machines in sustaining life led to intense 
anatomical investigations in the seventeenth century. Curiously, the 
emphasis shifted from organic machines during the eighteenth and 
nineteenth centuries; a critical analysis for plausible reasons has 
been carried out by Piccolino \cite{picco}. With the triumph of the 
molecular theory of matter in the beginning of the twentieth century, 
molecule became the prime object of investigation in chemistry as well 
as in biology. Interestingly, as early as in 1906, Loeb (as quoted in 
ref.\cite{taylor01}) referred to the cell as a ``chemical machine''.

Alice couldn't believe what she read. So, she asked her father, 
``does a cell really function as a machine?''. Her father replied,
``indeed, each of the cells in our body is like a `factory that 
contains an elaborate network of interlocking assembly lines' 
\cite{alberts98}, each of which is made up of an enormously large 
number of molecular machines. Just like macroscopic machines, these 
molecular machines also consist of `highly coordinated moving parts' 
\cite{alberts98,mavroidis,kinbara,schliwa,molloy,jphys,iyer,hackney,howard}. 
However, unlike man-made machines, these are products of Nature's 
{\it evolutionary design} over billions of years and the mechanisms 
underlying their function are often very different from their 
macroscopic counterparts in spite of many superficial similarities. 
The typical sizes of the molecular machines in our body cells range 
from tens to a few {\it nano-meters}. Therefore, one of the biggest 
challenges of modern science is to determine the principles behind 
the {\it design} of these {\it natural nano-machines} and to understand 
their dynamics which help in elucidating the {\it mechanisms} of their 
functions in terms of the fundamental principles of physics''. The 
aircraft now started running along the runway of the airport. Alice's 
father said, ''Alice, if you do not have a lot of home work during the 
next weekend, I'll be glad to tell you some of the intellectual 
challenges posed by the unusual mechanisms of these nano-machines 
inside our cells, each of which is like a micron-sized factory.

\section{A guided tour of the living cell in a dream} 

The aircraft lifted off the runway and was ascending. Alice looked out 
of the window and saw the busy vehicular traffic on the streets below. 
At one corner of the city she saw the high chimneys of the factories; 
that is the industrial area where her grandfather owned a small company 
that manufactured some chemicals. Few other high structures that she 
could recognise were huge water tanks that supplied drinking water to 
the adjoining residential areas. Several years ago she had seen one of 
the pumps which raised water up to those tanks from the water filtration 
plant below. 

Alice recalled the days when she was still in her kindergarten and her 
grandmother used to narrate the stories of Ramayana and Mahabharat in 
the evening during the summer vacations. The story she liked the most 
was that of the underwater kingdom of the snakes. She was so tired that,  
even before she realized, she was fast asleep in her seat and started 
dreaming.

Alice found herself at the entrance of the underwater kingdom. It was 
entirely covered by an approximately spherical membraneous soft wall.
A water molecule greeted Alice with a smile: ``Welcome Alice! Enjoy a 
guided tour of our kingdom which you call ``cell''. I'm Neera and I'll 
be your guide during the next ninety minutes of your guided tour of our 
kingdom''. 

Neera warned  Alice that the ``channel''-like passage of the entry gate 
was very  narrow and Alice would have to squeeze through it. The wall of 
the kingdom  actually consisted of two layers and the space in between 
was densely populated by what looked like two-tailed fishes; Neera seemed 
to have a special dislike for these fishes. When Alice asked the name of 
these unusual species of the fish, Neera said they are called {\it lipids}. 
The peculiar feature of these lipids was that the ones at the outer layer 
had their heads facing water outside the kingdom and those in the inner 
layer had their heads staring inside the kingdom; Neera said that the two 
monolayers of the lipids always stay together forming a bilayer. The 
tails of these fishes looked rather slippery and smelt like soap. 

While Alice was looking at the wall, spellbound by what she saw, Neera had 
donned a special swimming suit, somewhat like the ones used by swimmers 
at the Olympics. Neera said, I'll have tough time swimming across that 
channel unless I put on this swim suit. 

Just as they slipped in through the channel, Alice found a net of ropes 
which resembled fishing nets. First she thought that these were perhaps 
intended to prevent the two-tailed fishes from entering the kingdom. 
But, immediately she realized her mistake; she had never seen such a 
dual-purpose net. While taking off her swim suit, Neera said that this 
special net is called {\it cytoskeleton}. The girdirs and cables together 
form a skeleton that gives strength to the architecture of the kingdom  
and, at the same time, also form the back bone of the  transportation 
network of the kingdom \cite{howard,kreis}.

It was water everywhere inside the kingdom. But, unlike the freely 
flowing transparent water Alice had always been familiar with, it was 
rather viscous and turbid because of the enormous crowd swimming all 
around. Alice requested Neera for a brief pause as she wanted to watch 
the busy transportation system for a while before proceeding to their 
next destination.

%%%%%%%%%%%%%%%%%%%%%%%%%%%%%%%%%%%%%%%%%%%%%%%%%%%%%%%%%%%%%%
\begin{figure}[h]
\begin{center}
\includegraphics[angle=-90,width=0.5\columnwidth]{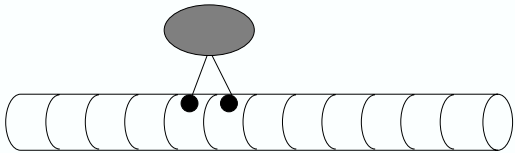}
\end{center}
\caption{A cartoon of a two-headed molecular motor moving on a filamentary 
track.
}
\label{fig-cytomot}
\end{figure}
%%%%%%%%%%%%%%%%%%%%%%%%%%%%%%%%%%%%%%%%%%%%%%%%%%%%%%%%%%%%%

Alice found that the motors were not moving smoothly; their jerky motion 
was similar to that of a person trying to move straight in a dust storm. 
The motors moving along the cables and struts had no wheels. They looked 
more like two-legged humans than the four-wheeled vehicles Alice 
was familiar with. Each carried a heavy bundle of cargo on their head and 
moved like porters. Alice's father owned a Toyota and her uncle had a Ford. 
Neera pointed at the motor passing by and said it was a {\it kinesin}. 
Another coming from the opposite direction was a {\it dynein}. Neera said 
that these two brands are like trains of the rapid mass transit system 
that ran from the city centre to the peripheral suburbs. The kinesins 
transport cargoes from the city centre to the peripheral subarbs while 
the dyneins move in the exactly opposite direction. The stiff struts 
that form the tracks for these motors are called {\it microtubules}. The 
shuttle buses that carried the passengers in these sububrbs were mostly 
{\it myosins} and they moved on more flexible cables called {\it actin}.

Alice did not find a single petrol pump (gas station) anywhere in the 
city and asked Neera how the motors were running without fuel. Neera  
corrected her mistake; the motors were consuming ``chemical fuels''. 
Each of the motors was utilizing the chemical energy  released by 
a chemical reaction. The fuels, namely molecules of a compound called 
ATP, were abundant everywhere and the motors were using the energy 
released by their hydrolysis to generate the mechanical energy required 
for their directed movement. Alice was well aware of the environmental 
problems created by the emissions from the motor vehicles back at home. 
So, she asked Neera how they disposed off the spent fuel. Neera said that 
all of the spent fuel is recycled in their kingdom; the spent fuel is 
recharged to manufacture fresh fuel. Neera also added that some other 
machines in their kingdom directly use sunlight and, therefore, are  
very eco-friendly as no waste product is generated by those machines. 
Moreover, another advantage of the machines running on light energy 
is that light can be switched on and off easily and rapidly.

%%%%%%%%%%%%%%%%%%%%%%%%%%%%%%%%%%%%%%%%%%%%%%%%%%%%%%%%%%%%%%
\begin{figure}[h]
\begin{center}
\includegraphics[angle=-90,width=0.5\columnwidth]{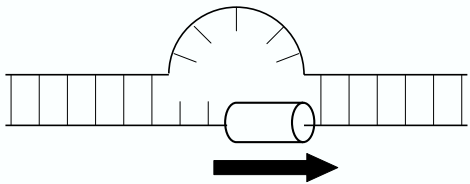}
\end{center}
\caption{A cartoon of a DNA helicase motor that moves on a nucleic 
acid track locally unzipping the double stranded DNA. 
}
\label{fig-helimot}
\end{figure}
%%%%%%%%%%%%%%%%%%%%%%%%%%%%%%%%%%%%%%%%%%%%%%%%%%%%%%%%%%%%%

\begin{figure}[h]
\begin{center}
\includegraphics[angle=-90,width=0.5\columnwidth]{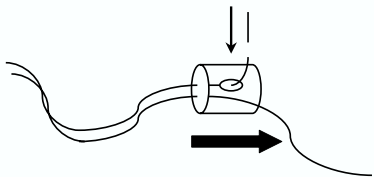}
\end{center}
\caption{A cartoon of a RNA polymerase motor that moves on a nucleic 
acid track synthesizing a m-RNA strand using a single-stranded DNA 
as the template. 
}
\label{fig-polymerase}
\end{figure}

As it was getting late, Alice resumed her sightseeing tour of the 
kingdom. Soon Alice and Neera were at the pore-like entrance to the 
palace, called {\it nucleus}, at the center of the kingdom. 
Alice was spellbound by the chemical factory inside the nucleus.
A member of a special class of machines, called DNA {\it helicase},   
was walking along a DNA fiber unwinding the two strands. 
This process resembled disentangling of two strands of fine thread 
that are  snarled together. The local opening of the double stranded 
DNA (see fig.\ref{fig-helimot}) is, effectively, an unzipping 
process that cuts the weak bonds which hold the two strands 
together. ``Such openings form one step in the overall processes 
of DNA repair or polymerisation of DNA and proteins'' explained 
Neera. RNA {\it polymerase}s (see fig.\ref{fig-polymerase}) also 
move on DNA and the main function of these machines is to polymerise 
the so-called messenger RNA joining nucleotides that are selected 
on the basis of the templates formed by a single-strand of DNA. 
The input energy for these machines comes from NTP condensation and 
the output is the work done by the machine against the opposing force.

%%%%%%%%%%%%%%%%%%%%%%%%%%%%%%%%%%%%%%%%%%%%%%%%%%%%%%%%%%%%%%
%%%%%%%%%%%%%%%%%%%%%%%%%%%%%%%%%%%%%%%%%%%%%%%%%%%%%%%%%%%%%

\begin{figure}[]
\begin{center}
\includegraphics[width=0.5\columnwidth]{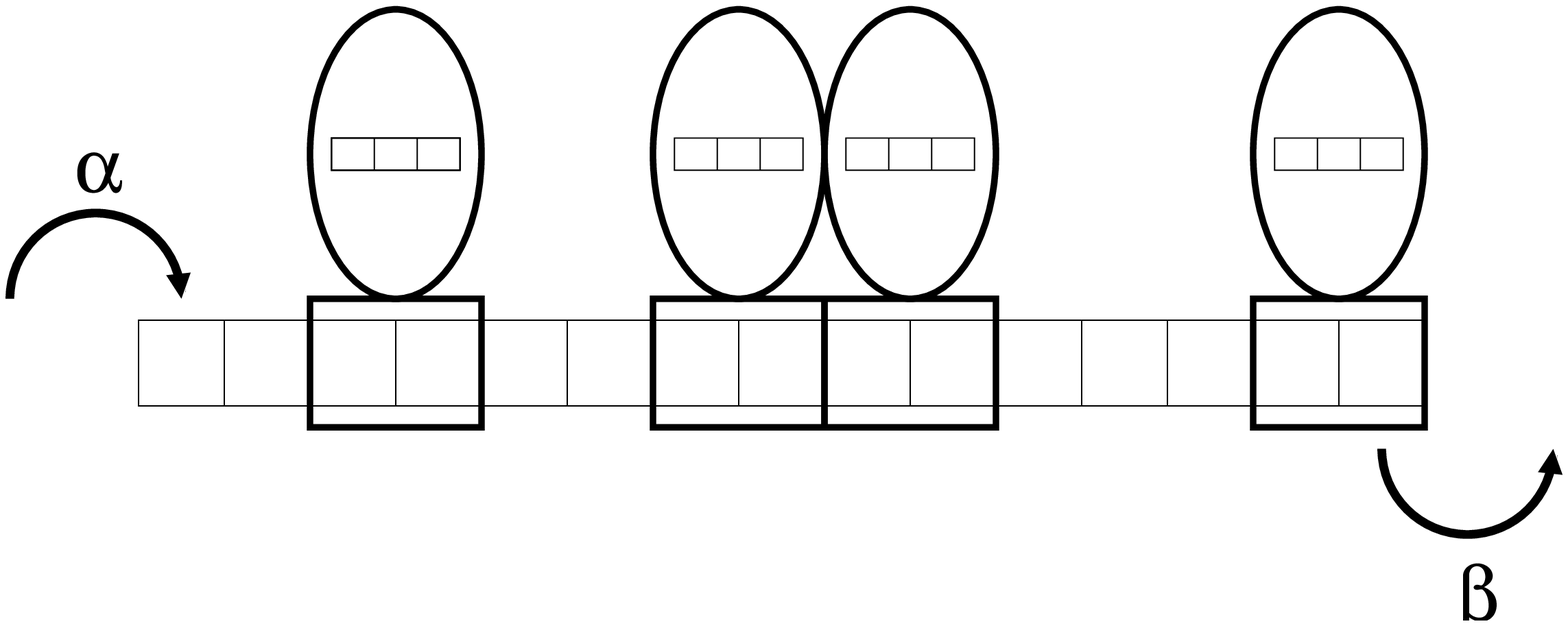}\\
\end{center}
\caption{A cartoon of a few ribosomes that move simultaneously on a 
m-RNA strand each synthesizing different copies of the same protein. 
The ribosome has two subunits; the larger subunit is represented 
schematically by the ellipse while the rectangular part represents 
the smaller subunit. The three small boxes on the larger subunit 
correspond to the three binding sites. The smaller subunit of the 
ribosome can cover simultaneously ${\ell}$ codons, i.e., ${\ell}$ 
triplets of nucleotides (${\ell} = 2$ in this figure) on the m-RNA.  
The parameters $\alpha$ and $\beta$ capture the effective rates of 
initiation and termination of translation.  
}
\label{fig-ribosome}
\end{figure}

Alice saw the synthesis of one such m-RNA strand in front of her eyes 
in a matter of seconds. Then, the serpentine m-RNA headed towards 
the nuclear pore. Neera said, ``after moving out of the nucleus, the 
m-RNA will serve as a template for the synthesis of a protein by 
another set of machines''. Alice was excited and wanted to see 
the whole process. She rushed out of the nucleus behind the m-RNA,  
closely followed by Neera. Alice saw that soon the m-RNA got 
decorated by little ball-like machines. She looked at Neera. ``Those 
are the {\it ribosomes}, explained Neera (see fig.\ref{fig-ribosome}). 
Alice saw a stiching-like action by each of the ribosomes, which also 
involved the m-RNA, t-RNA and amino acids, that, finally, gave birth 
to a fresh protein chain. However, almost instantaneously the nascent 
protein folded into a complex three-dimensional structure that Neera 
identified as the {\it tertiary} structure of the protein.

%%%%%%%%%%%%%%%%%%%%%%%%%%%%%%%%%%%%%%%%%%%%%%%%%%%%%%%%%%%%%%
%\begin{figure}[h]
%%%%%%%%%%%%%%%%%%%%%%%%%%%%%%%%%%%%%%%%%%%%%%%%%%%%%%%%%%%%%

\begin{figure}[h]
\begin{center}
\includegraphics[angle=-90,width=0.5\columnwidth]{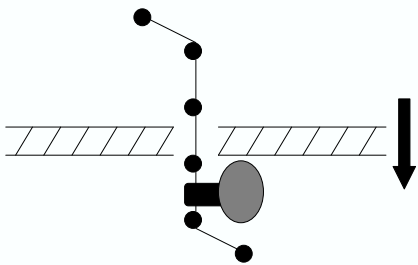}
\end{center}
\caption{A cartoon of a translocation machine that translocates a 
biological macromolecule across a membrane. 
}
\label{fig-trlocmot}
\end{figure}

On the way back, Alice saw many buildings of wide variety of shapes 
and sizes; each of these was enclosed by a soft membranous wall. 
Neera pointed her finger at one and said, ``can you see that 
labyrinthine structure? That's {\it endoplasmic reticulum}. And, of 
course, the small balls sticking to its membranous wall are the 
ribosomes.'' Then Neera turned in another direction, ``that one over 
there is called the {\it Golgi apparatus}. Alice could not control 
her curiosity. She asked, ``why do you need the Golgi apparatus?''. 
Neera was brief, ``The Golgi apparatus is like a letter sorting 
station of a postal system; the proteins are sorted here, packaged 
and forwarded to their onward destination.'' Neera added further, 
`` We'll visit some endoplasmic reticulum and Golgi apparatus during 
your future visits and show you the functions of those. In particular, 
there are some {\it translocation machines} on and near their walls 
which pull and push proteins in and out of the building. Without the 
help of these membrane-bound machines, no protein would be able to 
cross the walls of the endoplasmic reticulum and Golgi apparatus'' 
(see fig.\ref{fig-trlocmot}).

%%%%%%%%%%%%%%%%%%%%%%%%%%%%%%%%%%%%%%%%%%%%%%%%%%%%%%%%%%%%%%
%%%%%%%%%%%%%%%%%%%%%%%%%%%%%%%%%%%%%%%%%%%%%%%%%%%%%%%%%%%%%

%%%%%%%%%%%%%%%%%%%%%%%%%%%%%%%%%%%%%%%%%%%%%%%%%%%%%%%%%%%%%%
\begin{figure}[h]
\begin{center}
\includegraphics[angle=-90,width=0.5\columnwidth]{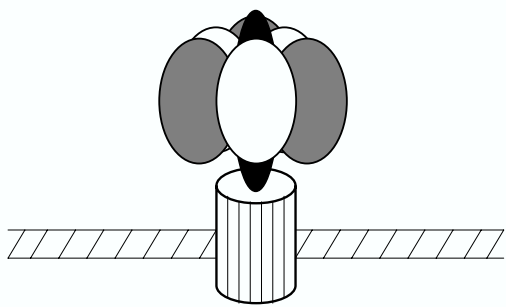}
\end{center}
\caption{A cartoon of ATP synthase, a rotary motor associated with 
a membrane. The cylindrical part is called the $F_0$ motor while 
the other part represents the $F_1$ motor. Both $F_0$ and $F_1$ 
are reversible and are tightly coupled to each other.
}
\label{fig-atpsynth}
\end{figure}
%%%%%%%%%%%%%%%%%%%%%%%%%%%%%%%%%%%%%%%%%%%%%%%%%%%%%%%%%%%%%

\begin{figure}[h]
\begin{center}
\includegraphics[angle=-90,width=0.5\columnwidth]{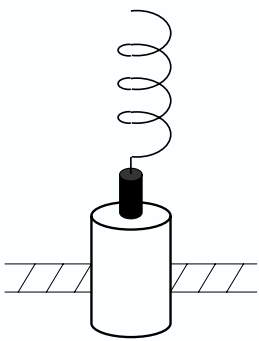}
\end{center}
\caption{A cartoon of the flagellar motor of bacteria, a rotary 
motor associated with a membrane. 
}
\label{fig-flagmot}
\end{figure}

Suddenly, Neera seemed very excited: ``Alice, can you see that small 
sausage-shaped building over there? That's a {\it mitochondrion}, one 
of the power houses that convert the spent fuel ADP into fresh fuel 
ATP.  It looks very different from the chemical factories in your 
world. This is achieved by a tiny machine, called ATP {\it synthase} 
\cite{oster}, stuck on the wall of the mitochondrion. Perhaps you 
cannot see it from here (see fig.\ref{fig-atpsynth}). But, I can 
assure you that this rotary motor currently holds the world record: 
it is the smallest rotary motor. This is, at least superficially, 
very similar to the motor of a hair dryer''. ''Is this the only 
rotary motor that exists in this kingdom?'', asked Alice. ``No. 
Bacteria also have a slightly larger rotary motor fixed on their skin 
\cite{berg}. This motor rotates a helical filament called a flagellum'' 
(see fig.\ref{fig-flagmot}).

%%%%%%%%%%%%%%%%%%%%%%%%%%%%%%%%%%%%%%%%%%%%%%%%%%%%%%%%%%%%%%
%%%%%%%%%%%%%%%%%%%%%%%%%%%%%%%%%%%%%%%%%%%%%%%%%%%%%%%%%%%%%

%%%%%%%%%%%%%%%%%%%%%%%%%%%%%%%%%%%%%%%%%%%%%%%%%%%%%%%%%%%%%%
\begin{figure}[h]
\begin{center}
\includegraphics[angle=-90,width=0.5\columnwidth]{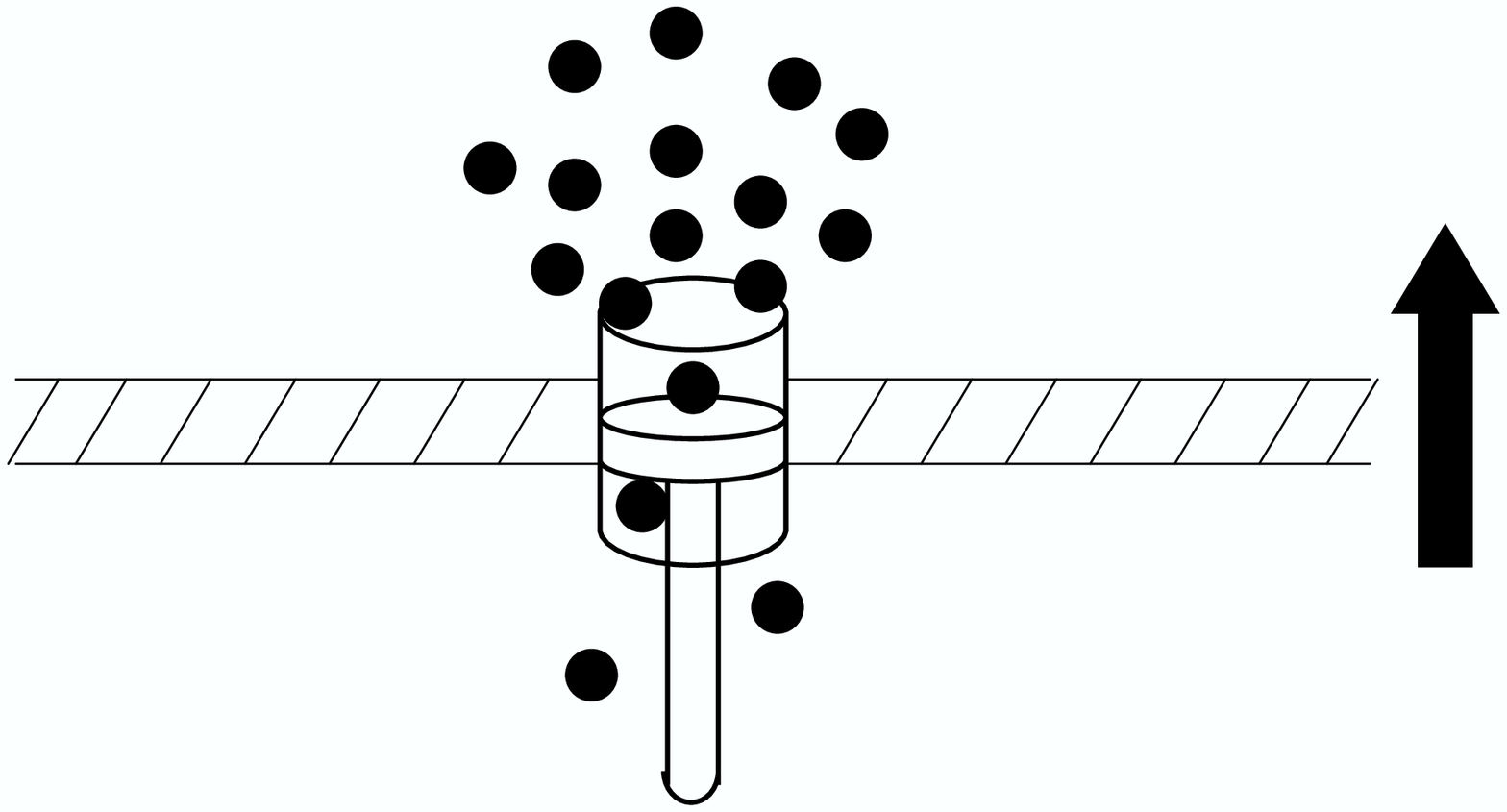}
\end{center}
\caption{A cartoon of a molecular ion pump associated with a membrane. 
}
\label{fig-ionpump}
\end{figure}
%%%%%%%%%%%%%%%%%%%%%%%%%%%%%%%%%%%%%%%%%%%%%%%%%%%%%%%%%%%%%
                                                                                
On their way back, Neera followed a shorter route and reached a 
gate different from the one through which they had entered. Neera 
wanted to show Alice a pump in action \cite{andersen}. It was fitted 
on the boundary wall. It was constantly pumping sodium ions from 
inside to outside while simultaneously, it was pumping potassium ions 
in the reverse direction. Neera emphasized, `` the pump is continuously 
consuming energy to carry the ions against the concentration gradient, 
i.e., opposite to the direction of the spontaneous flow, just as a 
water pump consumes electrical energy to draw water upward against 
gravity''.  

Alice and Neera came out of the kingdom through another slippery 
channel close to the pump. Neera said to Alice, ``I hope you liked our 
kingdom and will visit us again''. Unfortunately, this kingdom of 
ours will not last long''. This came as a shock to Alice, ``why 
will this cell perish?'' Neera consoled her, ``That's the rule in 
the world of cells; each goes through a life cycle and ultimately 
ends with a split into two daughter cells. This one is no exception 
and it will split into two daughter kingdoms in a few days. I'll be 
your guide during your next tour inside one of those two daughter 
kingdoms. Next time I'll also show you how cells use their machines 
to move from one place to another \cite{bray}. Some cells ``swim'' 
in their aqueous environment while others ``crawl''; the piston-like 
action of some internal machines play a crucial role in cell crawling''. 

Just as Alice and Neera were about to part ways, she saw a torpedo-shaped 
creature. It seemed as if it was looking for some way to sneak into 
the kingdom. It had a helical propeller attached to its skin that 
was rotating at an astonishingly high angular speed. Neera stopped 
Alice, ``beware! that is a bacterium''. Alice immediately realized 
the rotating propeller must be the flagellum that Neera had mentioned 
earlier. Neera continued, ``this bacterium is trying to invade the 
kingdom''. Alice saw yet another object that looked like a mortar 
shell. ``Oh my God, look at that virus'', exclaimed Neera in a shaky 
voice, ''this is another type of potential invader. If it succeeds, 
it will hijack our motors to get a free ride to the center of the 
town and then enter into the nucleus. If it can seize control of the 
machines involved in the protein synthesis, the future of the kingdom 
is doomed''. Suddenly the virus spotted Alice and, she thought, it 
was approaching her aggressively. Alice shrieked in fear... and her 
eyes opened. Her mother put her palm on Alice's head. ``What happened 
Alice?'', she asked. ``Nothing mom, I just saw something in my dream''. 
There was an announcement, the aircraft was going to land soon in Delhi.

As anticipated, Alice had very little home work during the next weekend. 
So she sat down with her father Albert and her elder brother Alex to 
hear the stories of molecular motors.\\
%and narrated what she had seen in her 
%dream. Her father listened quietly and, then remarked, ``Alice, the 
%kingdom you are describing is very much like the micron-sized factory-like 
%environment inside the cells of our body. You can learn more about the 
%nano-machines inside the cells from the notes I have prepared during the 
%last one week''. 

%The next few sections are excerpts from the note book of Alice's father.
\section{Discussion on methods, materials and mechanisms}

\noindent Alice: Can you really see an individual molecular machine 
with the help of some equipment in your lab as clearly as I saw them 
in my dream?\\
Albert: Seeing is believing. However, you may be surprised to hear that 
until late in the twentieth century scientists could not see individual 
molecules although everybody believed in their existence. In fact, in 
those days, the existence of a molecule could only be inferred indirectly 
from some circumstantial evidences gathered from experiments on samples 
which used to be very large compared to single molecules. Search for 
microscopes which would allow us to see this elusive object seemed like 
a ``holy grail''. We got our first glimpse of the macromolecules via 
X-ray diffraction and, then, electron microscopy. \\
Alice: Are the basic underlying principles of electron microscopy 
different from those of optical microscopy? \\
Albert: No. The basic principles of these two techniques are essentially 
similar. Optical microscopy uses light waves whereas the wave nature of 
electron is utilized in electron microscopy. The main difference is  
that much higher spatial resolution is achievable by electron microscopy 
because of the smaller wavelength of the electrons. \\
Alex: Why were scientists not satisfied even with the structures 
determined by electron microscopy?\\
Albert: What one got from those probes were static pictures. Towards 
the end of the twentieth-century a series of novel imaging techniques 
emerged which revolutionized optical microcopy. With the help of 
these tools we can now monitor the dynamics of single-molecules 
\cite{ajp}. Confocal fluorescence microscopy is, perhaps, the most 
versatile among these methods. The confocal microscope focusses on 
the intended plane of a sample and filters out the light from all the 
other planes.

%%%%%%%%%%%%%%%%%%%%%%%%%%%%%%%%%%%%%%%%%%%%%%%%%%%%%%%%%%%%%%
\begin{figure}[h]
\begin{center}
\includegraphics[angle=-90,width=0.5\columnwidth]{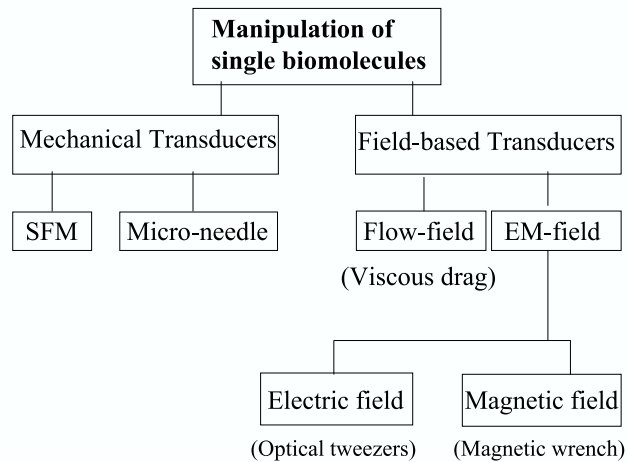}
\end{center}
\caption{A broad classification of the experimental techniques 
available for the manipulation of single molecules. 
}
\label{fig-manipul}
\end{figure}
%%%%%%%%%%%%%%%%%%%%%%%%%%%%%%%%%%%%%%%%%%%%%%%%%%%%%%%%%%%%%
 
\noindent Alex: Are these extraordinary developments in experimental 
techniques mainly responsible for the recent revolutions in cell  
biology?\\ 
Albert: Yes, indeed. After all, invention of novel experimental 
techniques always lead to enormous progress in natural sciences. 
For example, inventions of the telescope, microscope, X-ray opened 
up new horizons. In order to understand the mechanisms of molecular 
machines we need experimental probes with sufficiently high 
{\it spatial} as well as  {\it temporal} resolutions that would be 
adequate to watch the dynamics of these machines. Such high resolutions 
have been attained only in the last few years.\\
Alice: What are these new experimental rechniques for watching 
single molecules?\\
Albert: The recently developed single-molecule techniques can be broadly 
classified into two groups: (i) methods of imaging, and (ii) methods 
of manipulation. For example, green (and red) fluorescence microscopy 
enables us to monitor the dynamics of single motors just as ecologists 
use ``radio collars'' to track individual animals. But, for a clear 
understanding of the mechanism of the molecular machines it is not 
enough merely to watch them passively; we must also be able to 
manipulate them (see fig.\ref{fig-manipul}). Such manipulations have 
been possible over the last decade because of the availability of 
sophisticated techniques like, for example, optical tweezers and atomic   
force microscopes (AFM). \\  
Alice: Is there any similarity between ``optical tweezer'' and the 
tweezer grandma uses for pulling out grandpa's grey hairs?\\
Albert: Although the two look very different, their functions are 
very similar. In optical tweezers, oppositely moving laser beams 
create an optical trap using the radiation pressure of the laser 
light; this trap can grab a dielectric object which then can be 
manipulated just like what your grandma does with her tweezer.\\
Alex: What is the working principle of AFM?\\
Albert: The scanning force microscope (SFM) was originally designed to 
image a surface by scanning it with a sharp tip just like the finger 
tip of a human gives a fair picture of the macroscopic irregularities 
and the overall topography of a surface. In the AFM the tip is replaced 
by a cantilever that can pull, for example, a motor protein.\\
Alice: I thought only biologists study whatever goes on inside a cell. 
I didn't know that physicists like you are also working on these 
phenomena.\\
Albert: That, indeed, was true till a couple of decades ago. But, now 
this is a frontier area of interdisciplinary research that involves 
molecular cell biology, (bio-)chemistry, physics as well as engineering, 
especially, nano-technology \cite{frey}.\\
Alice: What is nano-technology? \\
Albert: Technologies which suitably utilize structures and processes 
that occur at the nanometer scale are collectively referred to as 
``nanotechnology''. For developing nanotechnology it is not enough 
just to get a fundamental understanding of structures and dynamics of 
materials at nanometer scale but one requires techniques to manipulate  
matter on such short scales so that novel applications can be 
innovated. \\
Alex: Is nano-technology a very recent development?\\
Albert: Although quest for ever-smaller size of devices began long ago, 
Feynman's famous talk of 1959 \cite{feynman} is accepted by the majority 
of physicists as the defining moment of nano-technology. Feynman first 
attracted the attention of the physics communiy to the unlimited 
possibilities of manipulating and controlling things on the scale of 
nano-meters. He indicated the potential advantages of nano-technology. 
He also speculated on the possible methods. Nano-technology is the 
latest trend in the twenty-first century. \\

Suddenly Albert's mobile phone started ringing; it was a call from one 
of his colleagues who, like Albert, switched to biological physics in 
the middle of his career a few years ago. The discussion resumed as 
soon as Albert's telephonic conversation ended. \\
Alex: What is the main goal of physicists working in this area of research?\\
Albert: Physicists use several different methods of investigation, namely, 
experimental, theoretical and computational techniques, for discovering 
the common fundamental physical principles that govern the generic 
features of the structure and dynamics of biological nano-machines.  
They also investigate the corresponding underlying mechanisms of 
operations of specific machines.\\ 
Alex: I understand the term ``experimental method'' and you have already 
given some examples of those techniques. But, I don't understand the 
term ``theoretical method''.\\
Albert: By ``theoretical methods'' we mean analytical (i.e., mathematical) 
treatment of theoretical models. Since such an analysis can be accomplished 
exactly only in rare cases, one has to make sensible approximations 
so as to get quantitative results as accurately as possible. \\
Alice: My biology teacher told us that the fruit fly {\it Drosophila} is 
a ``model organism''. Is there any relation between the terms ``model 
organism'' and ``theoretical model''?\\
Albert: No. In biology, often the simplest among a family of organisms  
is called a model system for the purpose of experimental investigations. 
For example, the fruit fly {\it Drosophila} and the worm {\it C-elegans} 
are often called model organisms because of their simpler features among 
all multi-cellular organisms. However, a theoretical model is an 
abstract representation of the real system. Biologists often develop 
{\it qualitative} models to explain empirical observations. In contrast, 
physicists work mostly with {\it quantitative} models, formulated in terms of 
mathematical symbols, not only to interpret experimental data but also  
to make new predictions that can be tested through experiments. \\
Alex: Wouldn't it be wise to develop an all encompassing single theoretical 
model that can explain all the intracellular phenomena?\\
Albert: Every theoretical model is intended to address a set of questions. 
The modeler must choose a {\it level of description} appropriate for 
this purpose keeping in mind the phenomena that are the subject of the 
investigation. Otherwise, the model may have either too much redundant 
details or it may be too coarse to provide any useful insight. For 
example, a molecular model of a hair dryer will have too much redundant 
information. Similarly, a continuum model of liquid water will be 
too coarse to study the dynamics of individual molecules of water. 
Since physicists most often focus only on generic features of the 
various classes of machines, rather than specific features of individual 
members of these classes, they normally develop minimal models which may 
be regarded as {\it mesoscopic}, rather than molecular, i.e., their status 
is somewhere in between those of the macroscopic and molecular models. \\
Alex: So far as the computational methods are concerned, are these 
identical to those used, for example, in bioinformatics?\\
Albert: Computational biology has two distinct branches- \\
(i) {\bf Knowledge discovery} (or, {\it data mining}) which extracts
hidden patterns or laws from huge quantities of experimental data,
forming hypotheses. Knowledge discovery is used extensively in 
bio-informatics. \\
(ii) {\bf Simulation-based analysis}, which tests hypotheses with 
computer simulations, provides predictions that, at least in principle, 
can be tested by in-vitro and/or in-vivo experiments in the laboratory. 
The starting point of a computer simulation is to develop an
{\it algorithm}, which can be implemented numerically, for
calculating the desired quantities. The sequence of instructions to
a computer specifying the numerical procedure of that algorithm
constitutes a computer {\it program}. \\
Alice: Are computer simulations different from what my biology teacher 
once referred to as {\it in-silico} experiments?\\
Albert: No, not at all. Computer simulation is often referred to as 
``computer experiments'' or, in analogy with in-vivo and in-vitro
experiments, also called {\it in-silico} experiments because of the 
close analogies between laboratory experiments and computer 
simulation. Laboratory experiments are performed on a sample of a 
material whereas computer simulation is an experiment with models.  
The computer program is the analogue of the experimentalist's apparatus, 
testing of a program with known and well understood models is the 
analogue of the calibration of the apparatus in laboratory. Computation 
is the analogue of experimental measurement and, finally, both laborary 
experiments and computer experiments require data analysis.\\
Alice: Can we simulate any arbitrary molecular machine with a computer?\\
Albert: No, there are practical limitations. The main difficulties faced 
in computer simulations arise from the limited size of the available 
computer memory and the limitations imposed by the available CPU time. \\

Alice's mother came into the room and placed the tea tray on the 
table. Albert started sipping tea while Alice's mother asked her, 
``Alice, are you getting satisfactory explanations of the things you 
saw in your dream?''. Alice smiled.\\
Alex: Since I have neither seen these machines in my dream nor 
observed them under the microscopes in your lab, could you please tell 
me what are the various types of molecular machines that operate inside 
an eukaryotic cell?\\
Albert:First, we can group the machines into two categories- one-shot 
machines and cyclic machines. The one-shot machines convert one form 
of energy (usually, chemical energy) into another (usually mechanical 
energy) only once, often in a sudden burst. In contrast, the machines 
belonging to the other group work in cycles just like engines of our 
automobiles. The cyclic machines, in turn, can be classified into 
different categories- motors, pumps, etc. The motors can also be 
divided into two categories, namely, linear and rotary; the linear 
motors move along filamentary tracks consuming fuel whereas rotary 
motors are like Alice's hair dryer.\\
Alex: Could you please give me examples of filamentary tracks used 
by the molecular motors?\\
Albert: Microtubules and flilamentary actin are most common tracks 
made of proteins. Nucleic acids like, for example, single-stranded 
DNA and mRNA serve as tracks for polymerases and ribosomes, 
respectively (see table \ref{tab-tracks} for typical examples of 
motors and the corresponding tracks).\\

%%%%%%%%%%%%%%%%%%%%%%%%%%%%%%%%%%%%%%%%%%%%%%%%%%%%%%%%%%%%%%%%
\begin{table}
\begin{tabular}{|c|c|} \hline
Motor  & Track   \\  \hline
Kinesin & Microtubule   \\  \hline
Dynein &  Microtubule   \\  \hline
Myosin & Actin filament \\  \hline
DNA helicase  & DNA strand  \\  \hline
RNA helicase  & RNA strand   \\  \hline
DNA polymerase & DNA strand  \\  \hline 
RNA polymerase & DNA strand  \\  \hline 
Ribosome       & RNA strand  \\ \hline 
\end{tabular}
\caption{Some common molecular motors and the corresponding tracks.
}
\label{tab-tracks}
\end{table}
%%%%%%%%%%%%%%%%%%%%%%%%%%%%%%%%%%%%%%%%%%%%%%%%%%%%%%%%%%%%%%%%
                                                                                
Alice: I have seen labourers laying down the railway tracks. Who 
plan and construct the network of the tracks inside the cells for 
the molecular motors?\\
Albert: In contrast to the railway tracks, the tracks for the 
molecular motors are dynamic.\\
Alice: Could you kindly give some examples of such dynamic tracks?\\
Albert: Microtubules, for example, are known to exhibit an unusual
polymerization-depolymerization dynamics even in the absence of motor
proteins. Moreover, in some circumstances, the motor proteins interact
with the microtubule tracks so as to influence their length as well as
shape; one such situation arises during cell division (the process is
called {\it mitosis}). Trains never create their track. But, a DNA 
helicase motor unwinds a double-stranded DNA and uses one of the single 
strands thus opened as the track for its own movement.\\
Alice: Wow! Do you identify the helicases, polymerases and ribosomes as  
motors just because they move on DNA or mRNA strands?\\
Albert: No doubt, they can be treated as molecular motors although, 
unlike cytoskeletal motors, they do not carry cargo. However, from 
a different perspective, a helicase is a ``unzipper'', a polymerase 
is a ``copying machine'' and the ribosomes are ``assembly lines''. 
In many circumstances, polymerizing microtubules and actin filaments 
act like nano-pistons. There are also examples of clamps and latches 
in eukaryotic cells.\\
Alex: Can one think of some other ways of classifying these molecular 
machines?\\
Albert: Sometimes the molecular machines are classified according to 
the environment where they operate. The cytoskeleton-based machines 
are constituents of the cytoskeleton or motors that move on cytoskeletal 
filaments. The helicases, polymerases, ribosomes, etc. are nucleic 
acid based motors whereas pumps, protein translocation machines, ATP 
synthase and flagellar motors are associated with membranes.\\ 
Alice: The machines which carry cargo are more like ``porters'' 
who carry luggage on their head while walking; unlike your car, these 
machines do not move on wheels.\\
Albert: You are right. First of all, none of the molecular motors in 
our cell has any part which even remotely resembles a wheel. In fact, 
there have been lively debates on this topic \cite{gould,dawkins} and 
it has been argued that in the soft world of living systems, it may 
be advantageous to have wheel-less transporters. Moreover, some authors 
have classified cytoskeletal motors into two groups: ``porters'' and 
``rowers''. Kinesins and dyneins are examples of cargo-carrying 
porters whereas myosins are typical examples of rowers.\\
Alice: Are you drawing any analogy between the myosins and the rowers 
whose rythmic movement of oars in and out of water impart high speed 
to the boats?\\
Albert: Yes, indeed. There are close similarities between the two 
situations. Each of the oars spend little time under water in each 
cycle, but the collective push of the rowers act constructively 
leading to the high speed of the boat. Similarly, each myosin spends 
only a small fraction of its biochemical cycle in contact with the 
actin track exerting a tiny force, but the collective action of a 
large number of the mysosin motors generates forces large enough 
for muscle contraction.\\

%%%%%%%%%%%%%%%%%%%%%%%%%%%%%%%%%%%%%%%%%%%%%%%%%%%%%%%%%%%%%%%%
\begin{table}
\begin{tabular}{|c|c|c|} \hline
Feature  & Macroscopic motors    & Molecular motors   \\  \hline
Material & Mostly Hard matter      & Mostly Soft matter   \\  \hline
Motor track & Road or rail      & Filamentary proteins or nucleic acids  \\  \hline
Fuel     & Mostly petrol or diesel & Mostly ATP or GTP   \\  \hline
Directionality  & Bidirectional & Unidirectional \\  \hline
Speed           & $\sim$ 1 Km/min. & $\sim$ 15 microns/sec  \\  \hline
Force           & $\sim$ 1000 N    & $\sim$ pN \\  \hline 
Energy          & 1-100 J     & 1-100 $\times 10^{-21}$ J \\ \hline 
\end{tabular}
\caption{Comparison of some of the features of macroscopic and molecular
machines.
}
\label{tab-comp}
\end{table}
%%%%%%%%%%%%%%%%%%%%%%%%%%%%%%%%%%%%%%%%%%%%%%%%%%%%%%%%%%%%%%%%

Alice was never interested in muscles. She was looking through the 
open window and saw their neighbour driving the car out of his  
garage in reverse gear. This triggered Alice's curiosity.\\
Alice: Isn't the mechanism of molecular machines very similar to those 
of their macroscopic counterparts? \\
Albert: Biomolecular machines operate in a domain where the appropriate 
units of length, time, force and energy are, {\it nano-meter}, 
{\it milli-second}, {\it pico-Newton} and $k_BT$, respectively 
($k_B$ being the Boltzamann constant). From a comparison of some of 
the characteristic features of the molecular motors and macroscopic 
motors (see table \ref{tab-comp}), naively, at first sight, 
one may think that the differences in 
the mechanisms of the molecular and macroscopic machines is merely a 
matter of two different scales (of size, time, force, energy etc.). But, that 
is not true. Since the masses of the molecular machines are extremely 
small, they are subjected to two dominating forces which are quite small 
for the macroscopic machines.\\
Alex: Does it mean that the molecular machines are not governed by the 
Newton's law that I have learnt in school?\\
Albert: Yes and no. The equation governing the motion of the machines 
is still Newton's equation, namely, mass $\times$ acceleration = force. 
But, there are crucial subtleties here. Because of its tiny mass, 
the inertial forces are orders of magnitude smaller than the viscous 
forces it experiences in the aqueous medium; in technical terms, one 
says that the dynamics of molecular motors is dominated by hydrodynamics 
at low Reynold's number \cite{purcell}. \\
Alice: What is the Reynold's number?\\
Albert: The Reynold's number ${\cal R}$ is the ratio of the inertial and 
viscous forces.  Let us estimate ${\cal R}$.  
It can be written as $R = \rho Lv/ \eta$ where $L$
is a characteristic linear size of the object, $\rho$ and $\eta$ are,
respectively, the density and viscosity of the liquid and $v$ is the
relative velocity of the object with respect to the liquid. For proteins
in aqueous solution,  $L \simeq 10 nm$, $\rho \simeq 10^3 Kg/m^3$,
$\eta \simeq 10^{-3} Pas$,  $v \simeq 1 m/s$ (corresponding to $1 nm/ns$).
Consequently, the corresponding Reynold's number is $R \simeq 10^{-2}$
(and even smaller for slower motions) 
whereas when you swim in our swimming pool ${\cal R}$ 
is of the order of $10^4$.\\
Alice: What difference does it make physically?\\
Albert: In order to appreciate the difficulties of swimming at low Reynold's
number, let us multiply $L$ by a factor of $10^7$, without altering
$\rho$ and v, so that $\eta$ must be multiplied by $10^7$ to keep
$R$ unaltered. In other words, if a motor tried to swim in an
aqueous medium adopting the swimming styles of humans, the difficulties
it would face will be very similar to those that you and I would face
if we ever tried to swim in honey.
Interestingly, because of the low Reynold's number, the nano-motors can 
come to a halt practically instantaneously as soon as they put on the 
brake.\\ 
Alex: Is that the only difference between molecular and macroscopic 
machines?\\
Albert: No. The molecular machines get bombarded from all sides by the 
randomly moving water molecules. Because of these bombardments, the 
machines experience an additional force which is totally random. 
The masses of molecular machines are so small 
that these {\it random thermal forces} have a strong infuence on their 
movement. It is because of these random forces that the molecular 
motors inside the cell move in a manner that resembles the movement 
of a human being in a violent  storm. \\ 
Alice: I do not have any feeling for the level of difficulty faced 
by the motors because of this storm-like situation. Could you 
kindly give us some concrete idea? \\
Albert: Let me consider sensory neurons, i.e., some special types 
of cells of your brain. If all the lengths are multiplied by $10^6$, 
then the transport from the cell center to the cell periphery is 
comparable to transportation of cargoes, which are supplied by a 
chemical plant of $10$ meter diameter, for more than $300$ kilometers 
along an approximately cylindrical pipe whose diameter is no more 
than $3$ meters. This cargo transport is to be achieved even when 
the transport vehicles are getting bombarded from all sides by a 
violent  storm!\\
Alice: What is meant by a ``random force''?\\
Albert: Well, even if you know the force acting on the machine at this 
instant of time, you have absolutely no idea of its magnitude and 
direction at any later time.\\
Alex: Then, how does one integrate the Newton's equation for the machine 
and find out its trajectory?\\
Albert: Very good question, Alex. In fact, the machine does not have 
any unique trajectory for a given initial condition; if you repeat the 
experiment, starting from identical initial conditions, the trajectories 
will be different from each other! In other words, the equation of 
motion for the molecular machine (for simplicity, in one dimensional 
space) is given by 
$$ 0 \times \frac{d^2x}{dt^2} = F_{load} ~~ - ~~ \gamma \frac{dx}{dt} ~~+ F_{br}$$ 
where the first term on the right hand side denotes the externally 
applied load force while the viscous drag force, captured through the 
second term on the right hand side, is assumed to be proportional 
to the velocity of the motor. The last term in this equation 
represents the random Brownian force.\\ 
Alex: Do you mean to say that the molecular motor behaves as a Brownian 
particle?\\
Albert: Yes, almost. However, unlike passive Brownian particles (like, 
for example, pollen grains in water), the molecular machines need  
(free-) energy input \cite{chowdhury} and, therefore, these are often 
regarded as ``active'' Brownian particles. At least some of these 
machines are physical realizations of Brownian ratchet, a device that 
exhibits directed, albeit noisy, movement in spite of being subjected 
to random Brownian forces as they operate in conditions far from 
thermodynamic equilibrium \cite{chowdhury,reimann,julicher}.\\
Alex: Is this relation with Brownian motion and non-equilibrium processes 
motivating a growing number of statistical physicists like you to get 
into this area of research?\\
Albert: These are certainly motivating many statistical physicists to 
work on biological machines. There are also other related motivating 
factors. The durable parts of macroscopic machines are normally ``hard'' 
so as to survive the regular wear and tear. In contrast, the structural 
elements of the cell, e.g., filaments and membranes, are ``soft''. These 
materials can be deformed easily because their conformations are 
determined by weak forces like, for example, Van der Waals forces, 
hydrogen bonding, etc. Moreover, since the thermal energy $k_BT$ 
available at a temperature $T$ is comparable to these weaker forces, 
thermal fluctuations can also lead to conformational changes. 
Furthermore, straightening of a filament or flattening of a membrane 
reduces its entropy and the corresponding restoring force is of entropic 
origin. Therefore,the free energies of soft matter are often dominated by 
entropy which gives rise to exotic phenomena that are not observed in 
hard materials. Naturally, not only statistical physicists like me, but 
also physicists who have been working for many years on soft matter are 
also getting interested in soft bio-materials.\\

%%%%%%%%%%%%%%%%%%%%%%%%%%%%%%%%%%%%%%%%%%%%%%%%%%%%%%%%%%%%%%
\begin{figure}[h]
\begin{center}
\includegraphics[angle=-90,width=0.5\columnwidth]{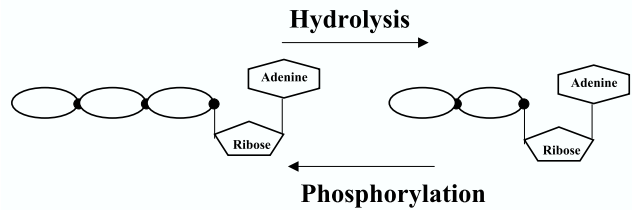}
\end{center}
\caption{Schematic representation of ATP and its hydrolysis.
}
\label{fig-atp2adp}
\end{figure}
%%%%%%%%%%%%%%%%%%%%%%%%%%%%%%%%%%%%%%%%%%%%%%%%%%%%%%%%%%%%%

The level of the discussion was now too high to be fully appreciated by 
Alice. She was watching the neighbour cleaning his car. She was 
thinking about the rising prices of petrol and cooking gas about which 
her mother had been complaining a lot in the last few days.\\
Alice: I could not understand how ATP molecules serve as fuels for 
molecular machines?\\
Albert: ATP is the energy currency of the cell and is composed of a 
sugar-base-phosphate combination found in nucleotides; the sugar 
of this energy currency is ribose, the base is adenine. However, ATP 
differs from RNA because, unlike RNA, it consists of more than one 
phosphate group. The adenosine triphosphate (ATP) has higher energy 
than adenosine diphosphate (ADP) and, the process whereby ATP gets 
converted to ADP and inorganic phosphate is called hydrolysis. The 
energy released by each ATP molecule through this process is about 
$10^{-19}$ J which corresponds to about $20 k_B T$ at room temperature. 
The reverse energy consuming process, through which the spent fuel is 
recharged, is called phosphorylation. Just as Euro is used in many 
places as the international currency, instead of US Dollar, some  
molecular machines use guanine triphosphate (GTP), instead of ATP, 
as the energy currency.\\ 
Alex: Does the presence of the track affect the rate of hydrolysis of 
ATP by the motor?\\

%%%%%%%%%%%%%%%%%%%%%%%%%%%%%%%%%%%%%%%%%%%%%%%%%%%%%%%%%%%%%%
\begin{figure}[h]
\begin{center}
\includegraphics[width=0.5\columnwidth]{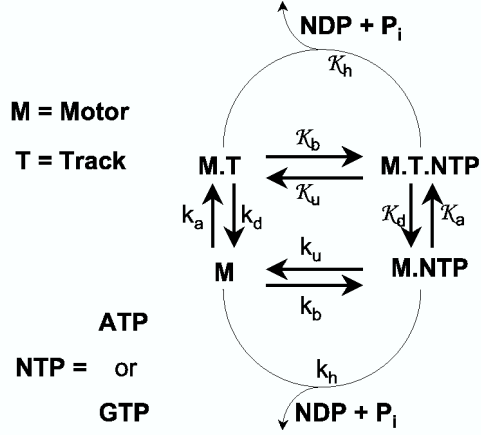}
\end{center}
\caption{Biochemical cycle in the absence and presence of the track. 
}
\label{fig-bccycle}
\end{figure}
%%%%%%%%%%%%%%%%%%%%%%%%%%%%%%%%%%%%%%%%%%%%%%%%%%%%%%%%%%%%%

Albert: Yes, indeed. The motor proteins work as enzymes even in the 
absence of the track. However, their enzymatic activity is boosted 
when the motor is bound to its track.\\

%%%%%%%%%%%%%%%%%%%%%%%%%%%%%%%%%%%%%%%%%%%%%%%%%%%%%%%%%%%%%%
\begin{figure}[h]
\begin{center}
\includegraphics[width=0.5\columnwidth]{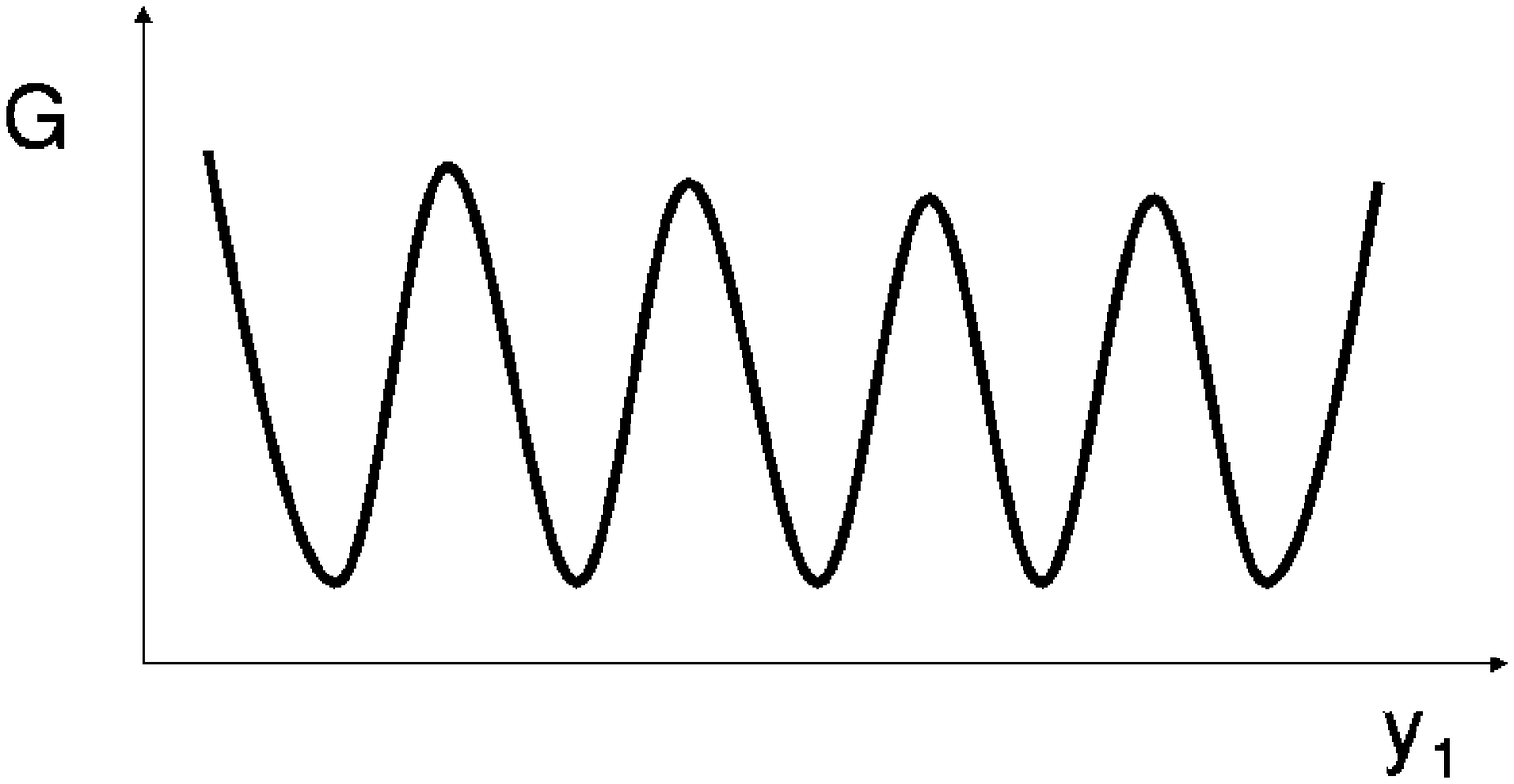}
\vspace{1cm}
(a)
\includegraphics[width=0.5\columnwidth]{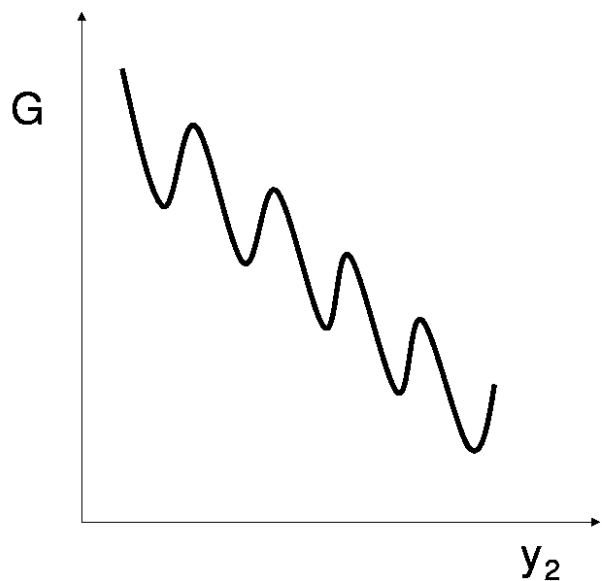}
\vspace{1cm}
(b)
\end{center}
\caption{Two different cross sections of the free energy landscape. 
Variation of the free energy $G$ with the ``mechanical variable'' 
$y_1$ and chemical variable $y_2$ are shown in (a) and (b), 
respectively. 
}
\label{fig-bccycle}
\end{figure}
%%%%%%%%%%%%%%%%%%%%%%%%%%%%%%%%%%%%%%%%%%%%%%%%%%%%%%%%%%%%%

Alex: Is this enhanced enzymatic activity of a motor in the presence 
of its track related to its walking ability along the track?\\
Albert: In the presence of the filamentary track, the mechanical and 
chemical cycles of the motor get coupled. For example, in case of 
conventional myosin, a mechanical step (the binding of the motor with 
the actin filament) catalyzes a chemical step (the release of phosphate) 
and, then, a chemical step (the binding of ATP) catalyzes a mechanical 
step (the detachment of the motor from the actin filament). Similarly, 
in case of conventional double-headed kinesin, a chemical step (the 
binding of ATP to head 1) catalyzes a mechanical step (the attachment 
of the head 2 to the MT) which, in turn, catalyzes a chemical step 
(the release of ADP from head 2) which, in turn, catalyzes a mechanical 
step (the detachment of head 1 from the MT) (see, for example, Howard 
2001). Thus, at the most abstract level, the problem of a molecular 
machine can be formulated as that of {\it mechano-chemistry} 
\cite{busta04}. \\
Alex: Could you kindly give me some intuitive physical picture of the 
{\it mechano-chemistry} of molecular motors?\\
Albert: Suppose, the system is being described by $y_1, y_2$, which 
represent the mechanical and chemical variables, respectively. The 
free energy $G(y_1,y_2)$ can be represented by a landscape such that 
$G(y_1,y_2)$ is the height (or, depth) of the landscape at the location 
$y_1,y_2$. In this scenario, the diffusion current along the mechanical 
coordinate is the velocity of the motor in real space, whereas that 
along the chemical coordinate is a measure of the enzymatic turnover 
rate. Thus, the landscape, i.e., the free-energy surface determines the 
kinetic mechanism of the motor \cite{keller1,keller2}. \\
Alex: I guess the landscape shows periodicity along $y_1$ because 
the positions of the binding sites for the motor are arranged 
periodically on its track. Am I right?\\
Alex: Yes, in case of cytoskeletal motors the perodicity of the MT and 
actin tracks give rise to periodicity of the landscape along $y_1$ 
for any arbitrary fixed value of $y_2$.\\
Alex: Is the landscape periodic also in the $y_2$ direction? \\
Albert: Along $y_2$, the free energy diagram looks exactly like a 
typical free energy diagram for a chemical reaction where the two 
minima corresponding to the reactants and products are separated by 
a free energy barrier. Since the enzyme returns to its initial state 
after each enzymatic cycle, $G$ is periodic also in $y_2$ except for 
an additional term that accounts for the free energy released by the 
reaction.  In other words,
\begin{eqnarray}
G(y_1+d,y_2) = G(y_1,y_2) \nonumber \\
G(y_1,y_2 +\Delta y_2) = G(y_1,y_2) + \Delta G 
\end{eqnarray}
where $d$ is the spatial period along $y_1$ (i.e., spacing between the 
successive binding sites on the track) while $\Delta y_2$ is the 
periodicity in the chemical variable.\\
Alice: Is it possible to visualize on the free-energy landscape the 
typical route of a motor?\\
Albert: Yes. In general, the landscape will have {\it local minima} 
connected by {\it low-energy passes}. These passes together define 
a low-energy path. The projection of the path on the $y_1-y_2$ plane 
is neither parallel to $y_1$ nor parallel to $y_2$. Consequently, 
the tilt of the landscape in the chemical direction drives the 
mechanical movement of the motor utilizing the free energy released 
the by chemical reaction to do the necessary mechanical work.\\ 

It was already lunch time and the family moved to the dining table. 
But the discussion continued.\\
Alex: What kind of fundamental questions are addressed in this area 
of interdisciplinary research?\\
Albert: Well, some of the fundamental questions are as follows:\\
(i) what are the {\it moving parts} of the
motor and what are their molecular constituents? \\
(ii) what is the {\it fuel} that supplies the (free-)energy
input for the machine ?\\
(iii) How does the {\it transduction} of energy take place? \\
(iv) What is the most appropriate definition of 
efficiency of the motor and how to estimate that efficiency?\\ 
(v) How is the operation of the machine {\it regulated} and 
{\it controlled}? For example, how is the machine switched on and off?\\
In addition to these general questions, several other fundamental 
questions are addressed in the context of specific molecular machines.\\
Alice: Is this research important from a biomedical perspective? Can 
the results of these investigations improve the quality of human life?\\
Albert: Yes. First, it is not surprising that defective molecular 
machines can cause diseases \cite{traffic,hirokawa} just as occasional 
disruption of work in any department of a factory can bring the entire 
factory to a standstill. In fact, 
malfunctioning of the track and/or motor can cause breakdown of the 
intracellular molecular motor transport system. Moreover, viruses 
are known to hijack the motors to travel from the cell periphery to 
the cell nucleus.  Therefore, if we understand how motors 
work at the molecular level we will not only be able to fix them when they 
malfunction but also control their progress and even find cures for 
the disease.  For example, we might also devise ways to selectively 
either arrest 
sub-cellular processes that cause diseases like cancer or slow down 
metabolism of organisms that invade cells causing other types of disease. 
The molecular motors can be used as vehicles for drug delivery \cite{cohen}. 
Thus, research on molecular machines can contribute towards improvement 
of  human health and fitness. \\
Alice: I am so impressed by mother nature!
I wish our factories were as sophisticated as the cell.\\ 
Albert: Alice, you may be glad to know that a large number of experts 
working in the area of nano-technology are drawing lessons from the 
principles used by nature in designing the molecular motors with the 
aim of synthesizing artificial nano-machines \cite{balzani}.\\
Alex: Are all the physicists and engineers following the same approach 
in nano-technology? \\
Albert: No.  The miniaturization of components for the fabrication of 
useful devices, which are essential for modern technology, is currently 
being pursued by engineers following mostly a top-down (from larger to 
smaller) approach. On the other hand, the natural molecular machines 
opened up a new frontier of nano-technology and provides an alternative 
approach- the bottom-up (from smaller to larger) approach \cite{balzani}. 
This approach is pursued mostly by chemists. Feynman, a great visionary, 
made the prophetic statement in his famous talk at the American Physical 
Society \cite{feynman}: ``ultimately, we can do chemical synthesis...
The chemist does a mysterious thing when he wants to make a molecule. 
He ....mixes this and that, and he takes it, and he fiddles around. And, 
at the end of a difficult process, he usually does succeed in synthesizing 
what he wants''. Indeed, in recent years, several artificial molecular 
machines have been synthesized chemically \cite{balzani}. In order to 
get some lessons from Nature's billion year experience in nano-technolgy, 
we must carry out reverse engineering. \\
Alice: What is reverse engineering?\\
Albert: Reverse engineering is ``the process of analyzing a system to 
identify the system's components and their interrelationships and create 
representations of the system in another form or at a higher level of 
abstraction \cite{chikofsky}. Reverse engineering of natural nanomachines 
will give us clues as to the possible design principles that can be 
utilized to synthesize artificial nanomachines. In fact, the term 
biomimetics has become popular buzzwords; this field deals with the 
design of artificial materials utilizing the principles of biomaterials 
\cite{barcohen}. \\

It was already late afternoon and Albert decided to revise a pending 
manuscript of a paper that needed minor amendments. He promised to 
narrate the story of each individual machine (i.e., their structure 
and function) during the next few weekends. But, before they dispersed, 
Albert asked Alex to summarize what emerged from their long discussion.\\ 
Alex: In recent years it has become clear that the cell is not a passive 
fluid medium. Rather, a cell is more like a factory where active 
biological processes are driven by coordinated operation of a large 
variety of molecular machines each of which performs a specific 
function. One of the challenges of fundamental research on these 
machines is to understand the mechanisms of their operation in 
terms of the physical principles governing their structure and 
dynamics. The results of these investigations are likely to have 
significant impact on applied research in biomedical sciences and 
have opened up novel approaches to nano-technology. Biomolecular 
nano-machines have become a subject of truly interdisciplinary 
research that involves biology, chemistry, physics as well as 
engineering and technology.  

\vspace{1cm}

\noindent {\bf Acknowledgements:} Indrani Chowdhury holds copyright 
of the figures; I thank her for giving me the required permission 
to reproduce the figures in this article. I also thank Aakash Basu 
and Debanjan Chowdhury for critical reading of the manuscript.


\begin{thebibliography}{99}    
\bibitem{picco} M. Piccolini, Nature Rev. Mol. Cell Biol. {\bf 1}, 149 (2000) 
\bibitem{taylor01} E.W. Taylor, Mol. Biol. of the cell {\bf 12}, 251 (2001).
\bibitem{alberts98} B. Alberts, Cell, 92, 291-294 (1998)
\bibitem{mavroidis} C. Mavroidis, A. Dubey and M.L. Yarmush, {\it Annu. Rev. Biomed. Eng}, (Annual Reviews, 2004).
\bibitem{kinbara} K. Kinbara and T. Aida, Chem. Rev. {\bf 105}, 1377 (2005).
\bibitem{schliwa} M. Schliwa, (ed.) {\it Molecular Motors}, (Wiley-VCH, 2003). 
\bibitem{molloy} J. E. Molloy and C. Veigel (eds.), Special issue of IEE Proceedings- Nanobiotechnology, {\bf 150}, No.3 (December, 2003).
\bibitem{jphys} Special issue of J. Phys. Cond. Matt. {\bf 17}, no.47 (2005). 
\bibitem{iyer} S. Iyer, B. Romanowicz and M. Laudon, A DARPA commissioned 
overview on ``Biomolecular Motors'' (with highlights from the special 
session at nanotech 2004, Boston, USA).
\bibitem{hackney} D. D. Hackney and F. Tanamoi, {\it The Enzymes}, vol.XXIII {\sl Energy Coupling and Molecular Motors} (Elsevier, 2004). 
\bibitem{howard} J. Howard, {\it Mechanics of Motor Proteins and the Cytoskeleton} (Sinauer Associates, massachusetts, 2001). 
\bibitem{kreis} T. Kreis and R. Vale, {\it Guidebook to the Cytoskeletal and Motor Proteins}, 2nd edition (Oxford University Press, 1999). 
\bibitem{oster} G. Oster and H. Wang, in ref.\cite{schliwa}. 
\bibitem{berg} H.C. Berg, {\it E. coli in Motion}, (Springer, 2003).
\bibitem{andersen} J. P. Andersen and E.E. Bittar, {\it Ion Pumps: Advances in molecular and cell biology}, vol.23A ( 1998).
\bibitem{bray} D. Bray, {\it Cell Movements: from molecules to motility} (Garland Publishig, Taylor and Francis, 2001).
\bibitem{ajp} J. M. Imhof and D.A. Vanden Bout, Am. J. Phys. {\bf 71}, 429 (2003).
\bibitem{frey} E. Frey, Chem.Phys.Chem. {\bf 3}, 270 (2002).
\bibitem{feynman} R. P. Feynman 1959, included also in: {\it The pleasure of finding things out}, (Perseus Books, Cambridge, Massachusetts, 1999), chapter 5.
\bibitem{gould} S. J. Gould,  {\it Hen's Teeth and Horse's Toes},(W.W. Norton 
and Co., 1984). 
\bibitem{dawkins} R. Dawkins, {\it The Sunday Times} Nov.24th (1996).
\bibitem{purcell} E.M. Purcell, Am. J. Phys. {\bf 45}, 3 (1977).
\bibitem{chowdhury} D. Chowdhury, Resonance {\bf 10(9)}, 63 (2005); {\bf 10(11)}, 42 (2005). 
\bibitem{reimann} P. Reimann, Phys. Rep. {\bf 361}, 57 (2002).
\bibitem{julicher} F. J\"ulicher, A. Ajdari and J. Prost, Rev. Mod. Phys. {\bf 69}, 1269 (1997).
\bibitem{busta04} C. Bustamante, Y.R. Chemla, N.R. Forde and D. Izhaky, Annu. Rev. Biochem. {\bf 73}, 705 (2004).
\bibitem{keller1} D. Keller and C. Bustamante, Biophys. J. {\bf 78}, 
541 (2000). 
\bibitem{keller2} C. Bustamante, D. Keller and G. Oster, Acc. Chem. Res. {\bf 34}, 412 (2001).
\bibitem{traffic} M. Aridor and L.A. Hannan, Traffic, {\bf 1}, 836 (2000); {\bf 3}, 781 (2002).
\bibitem{hirokawa} N. Hirokawa and R. Takemura, Trends in Biochem, Sci. {\bf 28}, 558 (2003).
\bibitem{cohen} R. N. Cohen, M.J. Rashkin, X. Wen and F. C. Szoka Jr., 
Drug Discovery Today: Technologies {\bf 2}, 111 (2005).
\bibitem{balzani} V. Balzani, M. Venturi and A. Credi, {\it Molecular devices and machines: a journey into the nano-world} (Wiley-VCH, 2003).
\bibitem{chikofsky} E.J. Chikofsky and J.H. Cross, II. IEEE Software 7, 13-17 (1990). 
\bibitem{barcohen} Y. Bar-Cohen, ed. {\it Biomimetics: biologically inspired technologies} (Taylor and Francies, 2005). 
\end{thebibliography}
\end{document}